\documentclass[aps,pra,twocolumn,superscriptaddress,nofootinbib]{revtex4-2}

\usepackage{graphicx}
\usepackage[dvipsnames]{xcolor}
\usepackage{amsmath}
\usepackage{amssymb}
\usepackage{bm}
\usepackage{hyperref}

\newcommand{\hb}{\hbar}
\newcommand{\Heff}{H_{\mathrm{eff}}}

\newcommand{\ii}{\mathrm{i}}
\newcommand{\new}[1]{{\color{black}#1}}

\begin{document}

\title{Effective Hamiltonians and non-Hermitian physics}

\author{Ugo Lair}
\affiliation{Laboratoire Collisions Agr\'egats R\'eactivit\'e,
UMR~5589, FERMI, Universit\'e de Toulouse, CNRS,
118~Route de Narbonne, 31062 Toulouse CEDEX~09, France}

\author{David Gu\'ery-Odelin}
\email{dgo@irsamc.ups-tlse.fr}
\affiliation{Laboratoire Collisions Agr\'egats R\'eactivit\'e,
UMR~5589, FERMI, Universit\'e de Toulouse, CNRS,
118~Route de Narbonne, 31062 Toulouse CEDEX~09, France}


\begin{abstract}
Every realistic quantum system is coupled to an environment.
Including the environment restores a closed, Hermitian
description, but at the cost of an intractably large Hilbert
space. The Feshbach--L\"owdin projection eliminates the
unwanted degrees of freedom exactly, yielding an effective
Hamiltonian $H_{\rm eff}$ on the relevant subspace alone---which
\new{generically becomes non-Hermitian when the eliminated
subspace contains an energetically accessible continuum and
outgoing boundary conditions are imposed}. We illustrate this through a minimal
model: a particle in a box with a delta barrier at its center.
Direct elimination of one half-well reduces all information
about the discarded sector to a single \new{boundary parameter} $\beta$; the same result is then rederived via the
Green's function and Dyson equation, offering a pedagogical
illustration of how boundary data enter the Green's-function
formalism and relating $\beta$
to a self-energy term.
Whether
$\beta$ is real or complex determines whether $H_{\rm eff}$ is
Hermitian or not, and the full range of behavior---from
Hermitian dynamics through finite-width resonances to the
counterintuitive quantum Zeno effect---follows from this single
parameter.
\end{abstract}

\maketitle

No laboratory system is truly isolated: an atom interacts
with vacuum fluctuations, a quantum dot is wired to metallic
leads, a spin couples to the phonon bath of its host crystal.
One can always insist on isolation by enlarging the system to
encompass the environment, but the resulting Hilbert space is then
intractably large. What one actually needs is an equation of
motion for the relevant degrees of freedom alone.

The Feshbach--L\"owdin projection
method~\cite{Feshbach1958,Feshbach1962,Lowdin1962,YaoShi2000,teVrugt2020}
provides exactly this. Introducing projectors $P$ onto the relevant
subspace $\mathcal{H}_L$ and $Q=1-P$ onto the eliminated
subspace $\mathcal{H}_R$, the full eigenvalue problem
$H|\psi\rangle=E|\psi\rangle$ reduces exactly to an effective
problem on $\mathcal{H}_L$:
\begin{equation}
  \Heff(E) = H_{LL} +
  \underbrace{H_{LR}\,(E - H_{RR})^{-1}\,H_{RL}}_{\text{self-energy }\Sigma(E)}.
  \label{eq:Heff_Feshbach}
\end{equation}
When the eliminated sector has only a discrete spectrum,
$\Sigma(E)$ is real on the real axis and $\Heff$ remains Hermitian.
\new{When the eliminated sector contains an open continuum---as for
a bath, a radiation field, or a semi-infinite lead---and the energy
lies within that continuum, the retarded self-energy generically
acquires an imaginary part, provided that the state under study is
actually coupled to the open channel; $\Heff$ is then
non-Hermitian, with complex poles encoding finite lifetimes. } Non-Hermitian physics is therefore
not an exotic modification of quantum mechanics: it is the
natural outcome of projecting an open system onto the
subspace one cares about.

Analytically transparent examples of this chain---from a
Hermitian Hamiltonian, through a Green's function, to an
explicit non-Hermitian $\Heff$---are relatively rare. The
purpose of this article is to exhibit one, using nothing
beyond matching conditions and a Wronskian, and to turn it
into a controlled setting in which the interplay between
boundary conditions and the Green's-function formalism
becomes fully transparent. Although our setting is deliberately
elementary, the boundary-response quantity that emerges from
eliminating the right sector is closely analogous in spirit
to the surface Green function of a lead, or equivalently to
an embedding self-energy in mesoscopic quantum transport:
the environment is reduced to an energy-dependent object
acting only at the contact \cite{Datta1995}.

The model is a particle of mass $m$ in a one-dimensional box
$[-a/2,a/2]$ with a repulsive delta barrier $+g\,\delta(x)$
($g>0$) at its center. On the left half-well $\mathcal{H}_L$,
$\Heff$ is the free kinetic operator, with all information
about the eliminated sector encoded in a single \new{boundary matching condition}
\begin{equation}
  \phi_L'(0) = \beta\,\phi_L(0).
  \label{eq:robin_intro}
\end{equation}
The coefficient $\beta$ is real for a closed
well with real $g$, purely imaginary for a purely imaginary
coupling, and necessarily complex for any real coupling in an
open geometry. The quantum Zeno effect~\cite{Misra1977,Facchi2002}
emerges as a direct consequence when $\beta$ is purely imaginary:
rather than draining the system faster, a stronger absorber
slows the decay and, in the limit of infinite absorption,
freezes the dynamics into a perfectly Hermitian one.

The article is organized as follows.
Section~\ref{sec:model} recalls the exact spectrum of the
model and identifies the even sector as the arena for all
non-trivial physics. Section~\ref{sec:elementary} derives
the \new{matching condition} by direct elimination of the right
half-well: this first approach is elementary but
model-specific, and already suffices to exhibit the quantum
Zeno effect in Sec.~\ref{sec:zeno}. 
To go further, Sec.~\ref{sec:green} rederives the same
results via the Green's function and Dyson equation;
presenting the two derivations in parallel turns the
Green's-function construction into a concrete algebraic
manipulation of the underlying wave function rather than an
abstract apparatus, recasts $\beta$ in an effective-boundary
language closely related to self-energy methods, and
provides a framework that does not rely on any closed-form
expression for the eliminated half.
The power of this reformulation is subsequently demonstrated
in Sec.~\ref{sec:tunneling}, where the closed right half-well
is replaced by a semi-infinite continuum: the effective boundary term acquires an imaginary part,
the \new{boundary parameter} becomes complex, and resonances of
finite width emerge as the inevitable consequence of an
open boundary. 

\section{The model and its exact spectrum}
\label{sec:model}

We consider a particle of mass $m$ in the symmetric well
$[-a/2, a/2]$ with a repulsive delta barrier at the center,
\begin{equation}
  H = -\frac{\hb^2}{2m}\frac{d^2}{dx^2} + g\,\delta(x),
\label{eq:H}
\end{equation}
with $g>0$ and $\psi(\pm a/2)=0$.
This ``particle-in-a-box plus delta'' model has a long
pedagogical tradition: it has been analyzed as a perturbation
problem~\cite{Lapidus1987}, through its strong- and
weak-coupling asymptotics~\cite{Joglekar2009}, by
factorization methods~\cite{PedramVahabi2010}, through a
variational treatment~\cite{Fernandez2021}, and more recently
with a detailed revisiting of its bound states under various
coupling signs~\cite{RamosdeCastro2024}. Our perspective here
differs: rather than solving the problem within the closed
box, we use it as the minimal arena in which projection,
effective Hamiltonian, and non-Hermiticity can be exhibited
in closed form. The delta barrier imposes
continuity of $\psi$ and a derivative jump at $x=0$,
\begin{equation}
  \phi'(0^+) - \phi'(0^-) = \frac{2mg}{\hb^2}\,\phi(0).
  \label{eq:jump}
\end{equation}
Because the barrier sits at the symmetry center, parity
$\mathcal{P}: x\mapsto -x$ commutes with $H$. Odd states
necessarily vanish at the origin and are completely immune to
the barrier; their energies are those of the empty well,
\begin{equation}
  E_n^{(-)} = \frac{2\hb^2\pi^2 n^2}{m a^2}, \quad n\geq 1.
  \label{eq:odd}
\end{equation}
This parity selection rule makes the model pedagogically
valuable: the Hilbert space cleanly separates into an
unperturbed sector (odd states) and an interacting sector
(even states), allowing the projection to be carried out
entirely within the even sector. For even states, the
left-half ansatz $\phi_L(x)=A\sin(k(x+a/2))/\sin(ka/2)$
with $k=\sqrt{2mE}/\hb$, normalized so that $\phi_L(0)=A$,
yields the exact quantization condition
\begin{equation}
  k\cot\!\left(\frac{ka}{2}\right) = -\frac{mg}{\hb^2}.
  \label{eq:quant_even}
\end{equation}
In the limit $g\to 0$, $k\to(2n-1)\pi/a$ (odd-indexed
levels of the empty well); for $g\to\infty$, $k\to 2n\pi/a$
(even-indexed levels of a half-well of length $a/2$). The
even-sector levels thus interpolate continuously between
these two limits, illustrating level repulsion.

\new{Let's introduce the single quantity around which the whole discussion will
revolve. On the left half-well the bulk equation is that of a
free particle, so an even eigenstate restricted to $[-a/2,0]$
is entirely determined by the Dirichlet condition
$\phi_L(-a/2)=0$ together with the value of its logarithmic
derivative at the barrier. We therefore introduce the
boundary parameter}
\begin{equation}
  \beta \equiv \frac{\phi_L'(0)}{\phi_L(0)},
  \label{eq:beta_def}
\end{equation}
\new{Once $\beta$ is known, the whole influence of the
barrier and of the right-hand sector on the left half-well is
encoded in the boundary matching condition
$\phi_L'(0)=\beta\,\phi_L(0)$: nothing else from $x\ge 0$ is
needed. Evaluating the left ansatz at the origin gives
$\beta=k\cot(ka/2)$, so that the exact quantization
condition~\eqref{eq:quant_even} is nothing but the statement
$\beta=-mg/\hb^2$. The parity selection rule ensures that
this single number carries all the non-trivial physics: the
odd sector, having $\phi(0)=0$, is blind to the barrier and
plays no role in what follows.}

\section{Effective Hamiltonian by direct elimination}
\label{sec:elementary}

We restrict attention to the left half-well $[-a/2,0]$ and
seek an effective operator $\Heff$ acting there, whose
eigenvalues coincide with those of $H$ in the even sector.
\new{This is the Feshbach--L\"owdin construction of
Eq.~\eqref{eq:Heff_Feshbach} in its most elementary
incarnation, with $P$ projecting onto $\mathcal{H}_L=L^2([-a/2,0])$
and $Q$ onto $\mathcal{H}_R=L^2([0,a/2])$; since the two halves
communicate only through the single point $x=0$, the boundary is
where the whole self-energy will end up, and $\beta$ is what
remains of it. $\phi_L$ and $\phi_R$ denote the restrictions of one
and the same even eigenstate to $[-a/2,0]$ and $[0,a/2]$
respectively; they are two distinct functions, each solving
a free Schr\"odinger equation in its own half-well, and they
are related only through the two matching relations
holding at $x=0$. We call interface the point
$x=0$ at which these two restrictions meet the barrier. The elimination now proceeds in three steps:
}

\new{\emph{Step (i): the eliminated (right) sector.}}
In the right half-well $[0,a/2]$, the solution satisfying
the Dirichlet wall at $x=a/2$ is
$\phi_R(x) = \phi_R(0)\,\sin(k(a/2-x))/\sin(ka/2)$.
\new{This sector is thus entirely characterized, as far as
the interface is concerned, by its own logarithmic
derivative}
\begin{equation}
  \frac{\phi_R'(0^+)}{\phi_R(0)} = -k\cot\!\left(\frac{ka}{2}\right).
  \label{eq:logderiv_R}
\end{equation}

\new{\emph{Step (ii): the two matching relations at the interface.}
The two sectors communicate only through continuity, which
transfers the amplitude, and Eq.~\eqref{eq:jump}, which
transfers the slope:}
 $ \phi_R(0)=\phi_L(0)\equiv A$ and $\phi_R'(0^+)-\phi_L'(0^-)=2mg\,A/\hb^2.$

\new{\emph{Step (iii): elimination of the right-hand data.}}
\new{Combining the previous equations and
isolating the left-hand data $\phi_L'(0^-)$ gives the energy-dependent boundary
matching condition}
\begin{equation}
  \phi_L'(0) = \beta_{\mathrm{eff}}(E)\,\phi_L(0),
  \qquad
  \beta_{\mathrm{eff}}(E)
  = -k\cot\!\left(\frac{ka}{2}\right)-\frac{2mg}{\hb^2}.
  \label{eq:beta_eff_closed}
\end{equation}
\new{The free solution in the
left half-well independently satisfies}
$  \frac{\phi_L'(0)}{\phi_L(0)} = k\cot\!\left(\frac{ka}{2}\right)$,
\new{and requiring this logarithmic derivative to coincide with the
boundary condition imposed by the eliminated sector,
$\beta=\beta_{\mathrm{eff}}(E)$, gives}
\begin{equation}
  k\cot\!\left(\frac{ka}{2}\right)
  = -k\cot\!\left(\frac{ka}{2}\right)-\frac{2mg}{\hb^2},
  \label{eq:selfconsistency}
\end{equation}
\new{that is, the exact quantization condition~\eqref{eq:quant_even}.
The eigenvalues of $H$ in the even sector are thus recovered as the
energies at which the effective boundary condition is compatible
with the left-hand dynamics---which is exactly what a
Feshbach--L\"owdin elimination is meant to deliver.}

\new{On shell, i.e.\ on a solution $E_n$ of
Eq.~\eqref{eq:selfconsistency}, the boundary parameter takes the
energy-independent value}
\begin{equation}
  \phi_L'(0) = \beta\,\phi_L(0),
  \qquad \beta \equiv \beta_{\mathrm{eff}}(E_n) = -\frac{mg}{\hb^2},
  \label{eq:robin}
\end{equation}
\new{with $\beta$ the boundary parameter of
Eq.~\eqref{eq:beta_def}. Off shell, the
boundary term generated by the elimination is energy dependent, as
Eq.~\eqref{eq:beta_eff_closed} makes explicit.}

The even eigenstates of $H$, restricted to $[-a/2,0]$,
therefore satisfy $\Heff\,\phi_L = E\,\phi_L$ with $x\in[-a/2,0]$,
where
\begin{equation}
  \Heff = -\frac{\hb^2}{2m}\,\frac{d^2}{dx^2},
\end{equation}
with boundary conditions $\phi_L(-a/2)=0$ and
Eq.~\eqref{eq:robin}. The bulk operator is the free
kinetic energy; all physics of the barrier and the
eliminated half-well is encoded solely in $\beta$.

Two features of this construction deserve emphasis.
First, \new{the on-shell value} $\beta=-mg/\hb^2$ is energy-independent\new{, whereas the
boundary term $\beta_{\mathrm{eff}}(E)$ produced by the elimination
is not}.
\new{The cancellation of the $k$-dependence occurs only on the
solutions of Eq.~\eqref{eq:selfconsistency}, and it occurs there
only because}
the delta sits exactly at the midpoint, making the left and
right propagators carry the same $k\cot(ka/2)$ factor.
For a delta at a generic position, or for a more
complex scatterer, \new{the on-shell value of} $\beta$ would depend on $E$ and the
eigenvalue problem would become nonlinear, requiring a
self-consistent solution\new{; the reflection symmetry of the present
model is what makes the answer a single number}. 
Second, the criterion for self-adjointness is entirely
controlled by $\mathrm{Im}(\beta)$: as shown in
Appendix~\ref{app:hermiticity}, the defect
$\langle\phi|\Heff\psi\rangle-\langle\Heff\phi|\psi\rangle$
reduces to a boundary term proportional to
$\mathrm{Im}(\beta)$, which vanishes for real $g$ and is
nonzero otherwise. Projecting a Hermitian system onto a
subsystem therefore need not break self-adjointness; the
outcome is determined solely by whether $\beta$ is real or
complex.

\section{Strong imaginary coupling: the quantum Zeno effect}
\label{sec:zeno}

\new{The previous section showed that $\Heff$ is self-adjoint if and
only if $\beta$ is real. What, then, does a complex $\beta$ describe?
The simplest way to produce one is to make the coupling itself
imaginary, $g\to-\ii\gamma$ with $\gamma>0$ (a sink, not a source),
so that the barrier becomes the anti-Hermitian potential
$-\ii\gamma\,\delta(x)$ in Eq.~\eqref{eq:H}. Such a term is an
absorber: a point-like channel at which probability leaves
the system and is not returned. It is no ad hoc device, but itself
the outcome of an elimination---the particle is coupled at $x=0$ to
degrees of freedom (photon modes of a resonant beam, detector
states, an external continuum) that are never observed again, and
eliminating them in the sense of Eq.~\eqref{eq:Heff_Feshbach} leaves
a self-energy which, for a fast and structureless (Markovian)
channel, reduces to a purely imaginary, energy-independent term at
the coupling point. As shown in Appendix~\ref{app:hermiticity},
$\Heff$ then generates the no-jump (null-measurement) evolution of a
monitored system, the missing norm being the probability that the
particle has been absorbed~\cite{BreuerPetruccione,Daley2014}.}
\new{For the full box with the imaginary delta potential, writing
$\ii\hb\,\partial_t\psi=(H_0-\ii\gamma\delta(x))\psi$ and
differentiating the total norm gives}
\begin{equation}
  \frac{d}{dt}\!\int_{-a/2}^{a/2} |\psi(x,t)|^2 dx
  = -\frac{2\gamma}{\hb}\,\bigl|\psi(0,t)\bigr|^2 .
  \label{eq:norm_loss}
\end{equation}
\new{The corresponding statement for the effective problem on the
left half-well alone follows from the boundary term computed in
Appendix~\ref{app:hermiticity}: with
$N_L=\int_{-a/2}^{0}|\psi_L|^2dx$,}
\begin{equation}
  \frac{dN_L}{dt}
  = -\frac{\hb}{m}\,\mathrm{Im}(\beta)\,\bigl|\psi_L(0,t)\bigr|^2
  = -\frac{\gamma}{\hb}\,\bigl|\psi_L(0,t)\bigr|^2 ,
  \label{eq:norm_loss_left}
\end{equation}
\new{the factor of two between the two expressions being simply
the second half-well, which loses probability at the same rate.
The probability is therefore lost only in proportion to
the density at the absorber.
An eigenstate of $\Heff$ has a
complex energy $E_n$ and evolves as
$e^{-\ii E_n t/\hb}$, so its norm decays exponentially,
$\|\psi_n(t)\|^2 = \|\psi_n(0)\|^2 e^{-\Gamma_n t}$, with the
\emph{decay rate} $ \Gamma_n \equiv -2\,\mathrm{Im}(E_n)/\hb \geq 0$, whose inverse is the lifetime of the mode. }

\subsection{\texorpdfstring{\new{The paradox: a stronger absorber decays more slowly}}{The paradox: a stronger absorber decays more slowly}}

We now consider a purely imaginary coupling $g = -\ii\gamma$ with
$\gamma>0$ (a pure absorber at $x=0$). The \new{boundary parameter}
becomes purely imaginary, $\beta = \ii m\gamma/\hbar^2$,
and the even-sector quantization condition reads
\begin{equation}
  k\cot\!\left(\frac{ka}{2}\right) = \frac{\ii m\gamma}{\hbar^2}.
  \label{eq:quant_zeno}
\end{equation}
All even eigenenergies are complex with
$\mathrm{Im}(E_n)<0$: every even mode decays. A naive
expectation is that a stronger absorber produces a larger
imaginary part; the exact solution shows the opposite.
A very strong absorber forces any eigenstate to have nearly
zero amplitude at $x=0$, but a state with exactly zero
amplitude cannot be absorbed.

\begin{figure}[t]
  \centering
  \includegraphics[width=\columnwidth]{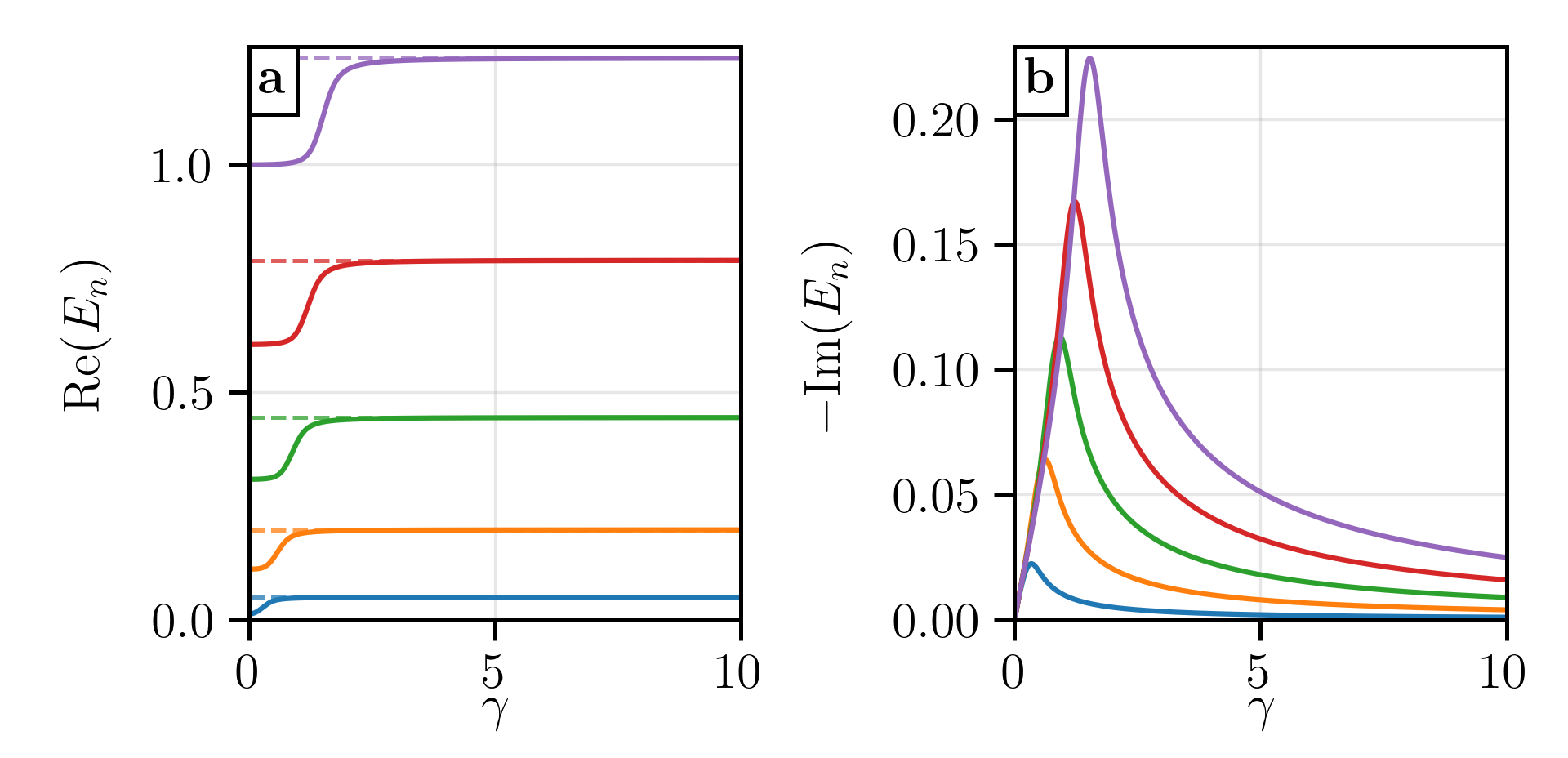}
  \caption{Complex spectrum of the even sector for a purely
    imaginary coupling $g=-\ii\gamma$ ($\gamma>0$), in units
    $\hbar^2/(ma^2)=1$, from Eq.~\eqref{eq:quant_zeno}.
    \textbf{(a)} Real part $\mathrm{Re}(E_n)$ as a function
    of $\gamma$ (solid curves); odd-sector energies
    $E_n^{(-)}=2\pi^2 n^2$ are shown as horizontal dashed
    lines. \textbf{(b)} \new{Resonance half-width}
    $-\mathrm{Im}(E_n)$ \new{(i.e.\ $\hb\Gamma_n/2$, with $\Gamma_n$
    the decay rate} as
    a function of $\gamma$, confirming the $1/\gamma$ Zeno
    suppression.}
  \label{fig:spectre}
\end{figure}
For large $\gamma$, writing $k_n = 2n\pi/a + \delta_n$ and
expanding gives $\delta_n \simeq -4n\pi\ii\hb^2/(m\gamma a^2)$,
so the decay rate of even mode $n$ is
\begin{equation}
  \Gamma_n = \frac{8\hb}{m\gamma a}\,E_n^{(-)}
  + O\!\left(\frac{1}{\gamma^2}\right).
  \label{eq:Gamma_zeno}
\end{equation}
This is the Zeno decay rate: $\Gamma_n\propto 1/\gamma$
decreases as the absorber strengthens.
Figure~\ref{fig:spectre} confirms this behavior: panel~(a)
shows the real parts migrating toward the nearest odd-sector
level as $\gamma$ grows, reflecting the effective Dirichlet
condition imposed by the strong absorber, while panel~(b)
confirms the $1/\gamma$ suppression.

As $\gamma\to\infty$, the \new{matching condition} transitions to
Dirichlet ($\phi_L(0)=0$), and $\Heff$ becomes the operator
of an isolated half-well: perfectly Hermitian. The
non-Hermitian dynamics self-projects onto the Hermitian Zeno
subspace---the space of states that avoid the absorber. This
is the quantum Zeno effect~\cite{Misra1977,Facchi2002}:
frequent measurement (here, continuous coupling to a loss
channel) confines the system to the dark subspace the
measurement cannot access.

\new{The condition $\phi(0)=0$ selects the domain of $\Heff$: as
$\gamma$ grows, that domain approaches the Dirichlet one at $x=0$,
and the long-lived dynamics is generated by the Hermitian
Hamiltonian of two disconnected half-wells ---the Zeno-subspace structure of Refs.~\cite{Facchi2002,FacchiPascazio2008}. Components
strongly coupled to the loss channel disappear during an initial
transient, but the widths of the survivors decrease as
$1/\gamma$: the law of Eq.~\eqref{eq:Gamma_zeno} is the competition
between the two factors of Eq.~\eqref{eq:norm_loss_left}, the loss
rate being the product of $\gamma$, which grows, and of
$|\phi(0)|^2$, which the same $\gamma$ suppresses as $1/\gamma^2$.
The absorber excludes probability from its own location until
nothing is left to absorb; the effect was observed on a trapped-ion
transition~\cite{Itano1990}.}

\begin{figure}[t]
  \centering
  \includegraphics[width=\columnwidth]{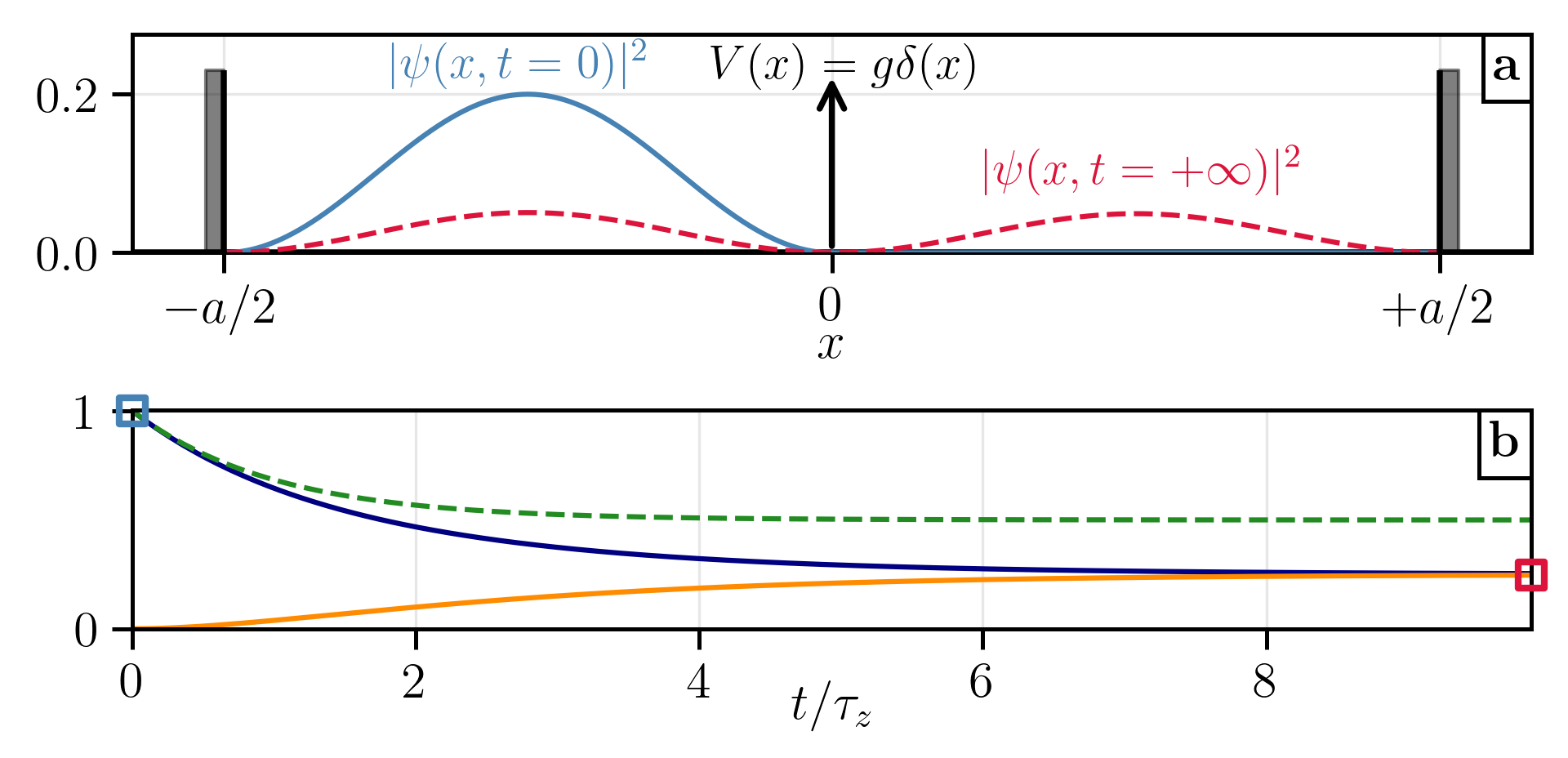}
  \caption{(Color online) Quantum Zeno dynamics, in units
    $\hbar^2/(ma^2)=1$.
    \textbf{(a)} Sketch of the system (box potential plus a \new{delta-function absorber at the center}). The initial and final \new{probability densities} are represented. For the initial state chosen here, the surviving state is the first odd mode
    $\phi_1^{(-)}(x)\propto\sin(2\pi x/a)$, with a node at
    $x=0$ and equal weight $1/4$ in each half-well. 
    \textbf{(b)} Left-well probability
    $P_L(t)=\int_{-a/2}^{0}|\psi(x,t)|^2\,dx$
    (solid blue line), right-well probability $P_R(t)$ (solid orange line) and $P_L(t)+P_R(t)$ (dashed green line)  as a function of rescaled time $t/\tau_Z$ .
    }
  \label{fig:probaandcavity}
\end{figure}

To illustrate the dynamics, we choose as initial state the
ground state of the isolated left half-well,
\begin{equation}
  \psi(x,0)=
  \begin{cases}
    \sqrt{\frac{4}{a}}\,
    \sin\!\left[\frac{2\pi}{a}\left(x+\frac{a}{2}\right)\right],
    & -\frac{a}{2}\le x\le 0,\\[1ex]
    0, & 0<x\le \frac{a}{2}.
  \end{cases}
\end{equation}
Because this state is entirely localized in the left half-well,
its odd and even parity components have equal norm. Moreover,
its odd component is exactly proportional to the first odd
eigenstate of the full box, $\phi^{(-)}_1(x)\propto \sin(2\pi x/a)$.
\new{In the strong-absorption regime, the long-time decay of the
even component is dominated by the lowest even resonance, hence
governed by} the Zeno timescale
$\tau_Z\equiv \Gamma_1^{-1}=m^2\gamma a^3/(16\pi^2\hbar^3)$,
whereas the odd component is immune to the absorber. Hence,
for $t\gg \tau_Z$, the dynamics is entirely governed by
$\phi^{(-)}_1$, and the long-time probability balance is
\begin{equation}
  \frac12+\frac14+\frac14=1,
\end{equation}
where the first term is the absorbed fraction and the last two
terms are the asymptotic probabilities in the left and right
half-wells (see Fig.~\ref{fig:probaandcavity}).
\new{Note that the two limits do not commute,}
  $\lim_{\gamma\to\infty}\lim_{t\to\infty}
  \;\neq\;
  \lim_{t\to\infty}\lim_{\gamma\to\infty}$
\new{For fixed finite $\gamma$, waiting until $t\gg\tau_Z$ removes
the even component and leaves the dark odd one, with the balance
above. At fixed observation time, $\gamma\to\infty$ sends
$\tau_Z\to\infty$ and gives unitary dynamics in the two
disconnected half-wells: the particle stays in the left well. Both
orders are physical---a fixed apparatus observed for a long time,
versus a freezing of the dynamics by an ever stronger
measurement.}

\section{The Green's function formalism}
\label{sec:green}

The derivation of Sec.~\ref{sec:elementary} relies on an
explicit closed-form expression for $\phi_R$. A more
general approach uses the Green's function
$G_R(0,0;E)$---the single quantity characterizing the
response of the right half at the \new{interface}---which carries
over unchanged when the right half is replaced by an
arbitrary environment \cite{Herrera2022,AtkinsonCrater1975,ErmanGadellaUncu2018}.

\new{The result of Sec.~\ref{sec:elementary} is already exact, and the
formalism that follows will, at first sight, only rederive it. Its advantage is modularity. The elimination of
Sec.~\ref{sec:elementary} used the explicit sine of the closed
right half-well and would have to be redone from scratch for any
other environment, whereas the Green's-function route stores the
eliminated sector in a single scalar function of the energy, its
boundary response $G_R^{-1}(E)$, and gives
$\beta_{\mathrm{eff}} \propto -g-G_R^{-1}(E)$. Replacing the closed
box by a lead, a bath, or a structured continuum then amounts to
substituting one formula for $G_R^{-1}$ by another, everything
else being unchanged---which is precisely what is done in
Sec.~\ref{sec:tunneling}. Two further benefits follow. First,
$G_R^{-1}(E)$ provides the boundary representation of the
embedding self-energy of the
Feshbach--L\"owdin construction, Eq.~\eqref{eq:Heff_Feshbach}:
because the spatial sectors of a differential operator communicate
through interface conditions rather than through an ordinary
matrix element, the self-energy appears here not as an additional
bulk potential but as an energy-dependent boundary term, so that
the correspondence with $H_{LR}(E-H_{RR})^{-1}H_{RL}$ is one of
role rather than of literal identity. Second, an environment with
a continuous spectrum makes $G_R^{-1}$ complex, so that
non-Hermiticity is read off directly from the analytic structure
of the response, without any need to postulate a complex
coupling. This mechanism underpins Fano resonances in quantum
transport~\cite{Fano1961,Miroshnichenko2010}, Anderson impurity
physics~\cite{Anderson1961,Hewson1993}, radiative linewidths in
cavity QED~\cite{CohenTannoudji1992,Haroche2006}, and
quasiparticle damping in metals~\cite{Mahan2000,Bruus2004}.}

\subsection{Green's function and Dyson equation}

For $E$ outside the spectrum of $H_0$, the resolvent kernel
$G_0(x,x';E)$ is defined by
$(E-H_0)G_0(x,x';E)=\delta(x-x')$.
It is meromorphic in $E$, with simple poles at the
eigenvalues of $H_0$, and on an interval with Dirichlet
walls admits the Wronskian representation
(see Appendix~\ref{app:wronskian})
\begin{equation}
  G_0(x,x';E)
  =
  \frac{2m}{\hb^2}\,
  \frac{u_1(x_<)\,u_2(x_>)}{W[u_1,u_2]},
  \label{eq:wronsk}
\end{equation}
where $u_1$ ($u_2$) vanishes at the left (right) wall.
Evaluating at $x=x'=0$ for the free box with $u_1(x)=\sin k(x+a/2)$ and $u_2(x)=\sin k(a/2-x)$, gives
\begin{equation}
  G_0(0,0;E)
  = -
  \frac{m}{\hb^2 k}\,
  \tan\!\left(\frac{ka}{2}\right).
  \label{eq:G0}
\end{equation}
This quantity diverges at the even levels of the empty well
(antinodes at $x=0$) and vanishes at the odd levels (nodes
at $x=0$)---a direct signature of the parity immunity noted
in Sec.~\ref{sec:model}.

For the Hamiltonian~\eqref{eq:H}, the full Green's function
satisfies the Dyson equation (see Appendix~\ref{app:dyson})
\begin{equation}
  G(x,x';E)
  =
  G_0(x,x';E)
  +
  g\,G_0(x,0;E)\,G(0,x';E).
\end{equation}
Because $V=g\,\delta(x)$ is a rank-1 perturbation, setting
$x=x'=0$ closes the equation algebraically:
\begin{equation}
  G(0,0;E)
  =
  \frac{G_0(0,0;E)}{1-g\,G_0(0,0;E)},
\end{equation}
the exact resummation of the Born series. The even-sector
eigenvalues are the poles of this expression, i.e.\ the
solutions of $G_0^{-1}(0,0;E) = g$, which reproduces
Eq.~\eqref{eq:quant_even} upon substituting Eq.~\eqref{eq:G0}.

\subsection{Effective boundary term and boundary matching condition}

To generalize beyond the closed well, we express
$G_0(0,0;E)$ in terms of the separate responses of the two
halves. Defining the logarithmic derivatives
$\Lambda_L \equiv u_L'(0)/u_L(0)$ and
$\Lambda_R \equiv -u_R'(0)/u_R(0)$ (with sign convention
following the outward normal from each half-well), the
symmetry $u_R(x)=u_L(-x)$ gives
$\Lambda_L = \Lambda_R = k\cot(ka/2)$.
Evaluating Eq.~\eqref{eq:wronsk} at the \new{interface} and factoring out $u_L(0)u_R(0)$ 
from the Wronskian gives $G_0 (0,0;E)=-(2m/\hb^2)/(\Lambda_L+\Lambda_R)$ so that, with $G_{L,R}^{-1}\equiv(\hb^2/2m)\Lambda_{L,R}$ one obtains
\begin{equation}
  G_0(0,0;E)
  = - \frac{1}{G_L^{-1}(E)+G_R^{-1}(E)},
  \label{eq:G0_DtN}
\end{equation}
where $G_{L,R}^{-1}(E) \equiv (\hb^2/2m)\,k\cot(ka/2)$.
This identity, which expresses the full \new{interface}
propagator as the series combination of two half-well
propagators, is the Green's-function counterpart of the
direct elimination performed in Sec.~\ref{sec:elementary}.
The secular equation $G_0^{-1}(0,0;E)=g$ then reads
$G_L^{-1}(E) + G_R^{-1}(E) = -g$.
Isolating the left half-well defines an effective boundary term acting on the left half-well,
\begin{equation}
  G_L^{-1}(E) = - g - G_R^{-1}(E) \equiv \Sigma_R(E).
  \label{eq:feshbach_cond}
\end{equation}
Using $G_L^{-1}=(\hb^2/2m)\phi_L'(0)/\phi_L(0)$ and
$G_R^{-1}=(\hb^2/2m)k\cot(ka/2) = - g/2$ on a solution of
the secular equation, one recovers directly $\beta=-mg/\hb^2$,
confirming that the two approaches are equivalent.

\new{The modularity announced at the beginning of this section is
now manifest in Eq.~\eqref{eq:feshbach_cond}: the barrier
contributes $-g$, the eliminated sector contributes
$-G_R^{-1}(E)$, and these are the only two ingredients of the
effective boundary term $\Sigma_R(E)$ felt by the left half-well.
Nothing has been gained in accuracy---the two routes are compared
line by line in Table~\ref{tab:compare}---but the right half-well
now sits in one clearly identified slot. If its content has a
continuous spectrum, $G_R^{-1}$ becomes complex, $\Sigma_R$
acquires an imaginary part, and $\Heff$ is non-Hermitian even for
real $g$: this is the subject of the next section.}

\section{Open boundary: a continuum as the discarded subspace}
\label{sec:tunneling}
We now remove the right wall while keeping the same projection strategy as in \new{Sec.~\ref{sec:green}}. The particle is still confined by an infinite wall at $x=-a/2$ and encounters the real barrier $g\,\delta(x)$ at the \new{interface} $x=0$, but the region $x>0$ is now an open continuum. We keep $\mathcal{H}_L=L^2([-a/2,0])$ as the subspace of interest \cite{Taron2013}, and continue to denote by $G_R$ the boundary response of the eliminated right sector.

In the present open geometry, the right sector is the free half-line $[0,+\infty)$ with outgoing boundary conditions. Its Dirichlet-to-Neumann map is simply
\begin{equation}
  \Lambda_R(E) \equiv -\frac{\phi_R'(0^+)}{\phi_R(0)} = -\ii k,
\end{equation}
so that, with the convention of \new{Sec.~\ref{sec:green}},
\begin{equation}
  G_R^{-1}(E)=\frac{\hb^2}{2m}\,\Lambda_R(E)
  =-\frac{\ii\hb^2 k}{2m}.
  \label{eq:GRinv_open}
\end{equation}
Unlike the closed half-well response, which is real and inherits the pole structure of a discrete spectrum, the open right sector contributes a purely imaginary boundary term: it represents an irreversible escape channel rather than a set of bound modes.

The boundary matching relation derived in \new{Sec.~\ref{sec:green}} remains unchanged,   $G_L^{-1}(E)+G_R^{-1}(E)=-g$.
Substituting Eq.~\eqref{eq:GRinv_open} gives
\begin{equation}
  G_L^{-1}(E)
  =-g+\frac{\ii\hb^2 k}{2m}.
  \label{eq:GLinv_open}
\end{equation}
Using $G_L^{-1}(E)=\hb^2\phi_L'(0)/(2m \phi_L(0))$,
we obtain the effective \new{matching condition} on the left half-well,
\begin{equation}
  \phi_L'(0)=\left(-\frac{2mg}{\hb^2}+\ii k\right)\phi_L(0).
  \label{eq:robin_tunnel}
\end{equation}
The real part $-2mg/\hb^2$ is the contribution of the delta barrier, whereas the imaginary term $+\ii k$ is the fingerprint of the continuum. Since $\mathrm{Im}\,\beta(E)=k\neq0$ for real $k>0$, the effective Hamiltonian on $[-a/2,0]$ is non-Hermitian for any real $g$: here non-Hermiticity is not a property of the coupling itself, but \new{the generic hallmark of an open boundary: it survives however weak the coupling, as long as the continuum is accessible}.
 Applying Eq.~\eqref{eq:robin_tunnel} to the left-well ansatz
$\phi_L(x)=A\sin\!\bigl(k(x+a/2)\bigr)$
yields the resonance condition
\begin{equation}
  k\left[\cot\!\left(\frac{ka}{2}\right)-\ii\right]
  =-\frac{2mg}{\hb^2}.
  \label{eq:quant_tunnel}
\end{equation}
Its complex roots $k_n$ determine the resonance energies
$E_n=\hb^2 k_n^2/(2m)$; their imaginary parts encode the leakage of probability from the left well into the continuum.
For real $g$, one finds $\mathrm{Im}(k_n)<0$, hence $\mathrm{Im}(E_n)<0$.
The decay rate \new{$\Gamma_n$ measures the probability loss into the continuum. An alternative derivation of \eqref{eq:quant_tunnel} is detailed in Appendix \ref{app:direct_open}}.

The coupling to the continuum also displaces the real part of $E_n$ away from the isolated half-well value, providing the reactive counterpart of the decay width. In this sense, the resonance pole carries both a dissipative component (the width) and a dispersive one (the frequency shift), much like the linewidth and level renormalization generated by a self-energy in open quantum systems. In more structured environments this dispersive contribution appears explicitly as the real part of the self-energy; in the present flat half-line model it is encoded implicitly in the position of the complex pole itself.

The origin of the decay is transparent at the level of the
boundary condition itself\new{: the positive imaginary part of
$\beta(E)$ in Eq.~\eqref{eq:robin_tunnel} is the signature of an
outgoing flux through the interface at $x=0$}.
Thus, the complex poles of Eq.~\eqref{eq:quant_tunnel}
represent genuine resonances: probability is not trapped in
the left well, but leaks irreversibly into the continuum.
This leakage vanishes in two physically distinct limits.
First, if the wavefunction amplitude at the \new{interface} tends
to zero, the state decouples from the continuum. Second,
near threshold $k\to0$, the group velocity in the continuum
vanishes, so escape is also suppressed. The first mechanism
is precisely the one at work in the Zeno regime discussed in
Sec.~\ref{sec:zeno}: a strong absorber expels probability
from the \new{interface}, driving $|\phi_n(0)|\to0$ and thereby
reducing, rather than enhancing, the decay.
\section{Summary and outlook}

A single scalar parameter---the \new{boundary parameter} $\beta$
defined by $\phi_L'(0)=\beta\,\phi_L(0)$---unifies four
physical regimes of the same model: Hermitian dynamics,
non-Hermitian decay by design, resonances induced by openness,
and the quantum Zeno effect. In every case the bulk operator
$\Heff=-(\hbar^2/2m)\partial_x^2$ is unchanged; all the
physics of the eliminated sector is encoded solely in $\beta$.
Two lessons stand out. First, projection reshapes boundary
conditions, not bulk dynamics. Second, non-Hermiticity
arising from openness \new{requires no fine-tuning: whenever the
discarded subspace contains an accessible continuum to which the
state is actually coupled, and outgoing conditions are imposed,
the boundary self-energy acquires an imaginary part at real
energies and $\Heff$ becomes non-Hermitian regardless of the
coupling strength}.

\new{According to Eq.~\eqref{eq:Heff_Feshbach}, eliminating the
$Q$-sector produces an effective Hamiltonian
$\Heff(E)=H_{LL}+\Sigma(E)$ on the $P$-sector alone, with all the
eliminated physics stored in the self-energy
$\Sigma(E)=H_{LR}(E-H_{RR})^{-1}H_{RL}$. Here every element of it
is explicit. The
$P$-sector is $L^2([-a/2,0])$ and the $Q$-sector is whatever
lives at $x\ge 0$; the coupling $H_{LR}$ is supported at the
single point $x=0$---so that the two sectors communicate through
interface conditions rather than through an ordinary matrix
element---which is why the eliminated sector does not
renormalize the bulk operator $H_{LL}=-(\hb^2/2m)\partial_x^2$ at
all, but leaves an energy-dependent boundary term; the resolvent
$(E-H_{RR})^{-1}$ of the eliminated sector becomes its boundary
response $G_R(E)$, computed in
Sec.~\ref{sec:green} for a closed half-well and in
Sec.~\ref{sec:tunneling} for a continuum; and the boundary
representation of the embedding self-energy is
$\Sigma_R(E)=-g-G_R^{-1}(E)$, which is $\beta_{\mathrm{eff}}(E)$ up
to the factor $\hb^2/2m$. Comparing Eq.~\eqref{eq:robin} with Eq.~\eqref{eq:robin_tunnel}
then exhibits the mechanism stated below
Eq.~\eqref{eq:Heff_Feshbach} in its simplest setting: the closed
half-well gives $\beta=-mg/\hb^2\in\mathbb{R}$, whereas the open
half-line adds the term $+\ii k$ inherited from the continuum,
whose imaginary part is the decay width. The delta barrier in a box
is thus a Feshbach--L\"owdin projection carried out by hand from
beginning to end.}

The two derivations of $\beta$---direct elimination and the 
Green's-function route---converge to the same result, and it 
is precisely the redundancy that is instructive. By 
deriving $\beta$ twice, in two languages, one sees each step 
of the Green's-function construction reflected in an explicit 
algebraic manipulation of the underlying wave function. 
\new{Table~\ref{tab:compare} summarizes the correspondence: the two
columns differ not in their content but in their extensibility.}

\begin{table*}[t]
  \caption{\new{Correspondence between the two derivations of the
    boundary parameter $\beta$.}}
  \label{tab:compare}
  \centering
  \begin{tabular}{p{0.22\textwidth} p{0.35\textwidth} p{0.35\textwidth}}
    \hline
    & Direct elimination (Sec.~\ref{sec:elementary})
    & Boundary response (Sec.~\ref{sec:green})\\
    \hline
    Object describing the eliminated sector
      & closed-form $\phi_R$, Eq.~\eqref{eq:logderiv_R}
      & response $G_R^{-1}(E)$, Eq.~\eqref{eq:G0_DtN}\\
    Tool used at $x=0$
      & continuity + jump condition
      & Dyson equation, rank-1 closure\\
    Result of the elimination
      & $\beta_{\mathrm{eff}}(E)=-2mg/\hb^2-k\cot(ka/2)$,
        Eq.~\eqref{eq:beta_eff_closed}
      & $\Sigma_R(E)=-g-G_R^{-1}(E)$, Eq.~\eqref{eq:feshbach_cond}\\
    On-shell value
      & $\beta=-mg/\hb^2$
      & $\beta=-mg/\hb^2$ (same)\\
    Prerequisite
      & the right sector must be solvable in closed form
      & only $G_R^{-1}(E)$ is needed\\
    Extension to a continuum
      & possible but must be redone (Appendix~\ref{app:direct_open})
      & substitute $G_R^{-1}=-\ii\hb^2k/2m$\\
    \hline
  \end{tabular}
\end{table*}

Here the elementary approach builds intuition\new{,} while the 
boundary-response formulation opens the door to arbitrary 
environments: a structured spectral density would introduce
memory effects and non-Markovian dynamics~\cite{BreuerPetruccione},
making $\beta$ frequency-dependent; promoting the model to a
many-body setting---for instance a Tonks--Girardeau gas in
the left half-well---would connect the \new{boundary-matching physics}
to correlated open quantum systems~\cite{Daley2014}.

\begin{acknowledgments}
D.G.-O.\ acknowledges support from the Institut Universitaire
de France.
\end{acknowledgments}

\section*{Author contributions}
D.G.-O. conceived the study and led the investigation. U.L. performed the formal analysis, validated the results, and critically revised the manuscript.

\section*{Data availability}
No new data were created or analysed in this study. All
results are reproducible from the equations provided in the
manuscript.

\appendix
\section{Self-adjointness and non-Hermiticity of $\Heff$}
\label{app:hermiticity}

For any two functions $\phi,\psi$ satisfying the same
Dirichlet condition at $x=-a/2$ and the same \new{matching condition}
$\phi'(0)=\beta\phi(0)$ at $x=0$, integration by parts gives
\begin{equation}
  \langle\phi|\Heff\psi\rangle - \langle\Heff\phi|\psi\rangle
  = -\frac{\ii\hb^2}{m}\,\mathrm{Im}(\beta)\,\phi^*(0) \psi(0),
\end{equation}
so that $\Heff$ is self-adjoint if and only if
$\mathrm{Im}(\beta)=0$. Real $g$ gives
$\beta=-mg/\hbar^2\in\mathbb{R}$, whereas $g=-\ii\gamma$
($\gamma>0$) gives $\mathrm{Im}(\beta)=m\gamma/\hbar^2$ and a
departure from self-adjointness entirely localized at $x=0$---as it
must be, the defect of any Sturm--Liouville operator being a
boundary term~\cite{Kato1966,BonneauFarautValent2001}.

The resulting term $-\ii\gamma|\phi(0)|^2$ is the precursor of the
Lindblad description of a local particle-loss channel:
\new{the imaginary potential generates the no-jump evolution alone,
the jumps being described by
$\hat L=\sqrt{2\gamma/\hb}\,\hat\psi(0)$, which removes the particle
rather than projecting it at $x=0$, and which reproduces the loss
term of Eq.~\eqref{eq:H} through $-(\ii\hb/2)\hat L^\dagger\hat L$.}

\section{Wronskian representation of the Green's function}
\label{app:wronskian}

Let $u_1,u_2$ be two linearly independent solutions of
$(E-H_0)u=0$ satisfying the left and right boundary
conditions respectively. For $x\neq x'$, the Green's
function must take the form $G=Au_1(x)$ for $x<x'$ and
$G=Bu_2(x)$ for $x>x'$. Continuity at $x'$ gives
$Au_1(x')=Bu_2(x')$, and integrating
$(E-H_0)G=\delta(x-x')$ across $x'$ gives the jump
condition $\partial_x G|_{x'^+}-\partial_x G|_{x'^-}=2m/\hb^2$.
Solving for $A$ and $B$ yields
\begin{equation}
  G_0(x,x';E)
  =
  \frac{2m}{\hb^2}\,
  \frac{u_1(x_<)\,u_2(x_>)}{W[u_1,u_2]},
\end{equation}
where $x_< = \min(x,x')$, $x_> = \max(x,x')$, and $W[u_1,u_2]=u_1u_2'-u_1'u_2$ is the (constant)
Wronskian.

\section{Dyson equation from first principles}
\label{app:dyson}

Let $G_0=(E-H_0)^{-1}$ and $G=(E-H_0-V)^{-1}$.
Writing $E-H_0=(E-H_0-V)+V$, multiplying left by $G_0$
and right by $G$ gives $G=G_0+G_0VG$. In position
representation with $V(y)=g\,\delta(y)$:
\begin{equation}
  G(x,x';E) = G_0(x,x';E) + g\,G_0(x,0;E)\,G(0,x';E). \nonumber
\end{equation}
Setting $x=x'=0$ closes the equation at the single point
$(0,0)$ and yields $G(0,0;E)=G_0(0,0;E)/[1-gG_0(0,0;E)]$,
the exact resummation of the Born series.


\section{Direct derivation of the open-boundary matching condition}
\label{app:direct_open}

For completeness, we briefly rederive the effective boundary condition of the open geometry without using Green's functions. In the right region $x>0$, the outgoing solution at energy $E=\hb^2 k^2/(2m)$ is
$\phi_R(x)=C\,e^{\ii kx}$.
Continuity at the \new{interface} gives $C=\phi_L(0)$, hence
$\phi_R'(0^+)=\ii k\,\phi_L(0)$.
Using the jump condition across the delta barrier, we obtain
\begin{equation}
  \phi_L'(0)=\left(\ii k-\frac{2mg}{\hb^2}\right)\phi_L(0),
\end{equation}
which is Eq.~\eqref{eq:robin_tunnel} of the main text. Finally, inserting the left-well form $\phi_L(x)=A\sin\!\bigl(k(x+a/2)\bigr)$
gives Eq.~\eqref{eq:quant_tunnel} independently of the Green-function formalism.



\begin{thebibliography}{99}

\bibitem{Feshbach1958}
  H.~Feshbach, Ann.\ Phys.\ \textbf{5}, 357 (1958).

\bibitem{Feshbach1962}
  H.~Feshbach, Ann.\ Phys.\ \textbf{19}, 287 (1962).

\bibitem{Lowdin1962}
  P.-O. L\"{o}wdin, J.~Math.\ Phys.\ \textbf{3}, 969 (1962).

\bibitem{YaoShi2000}
  D.~Yao and J.~Shi,
  ``Projection operator approach to time-independent
  perturbation theory in quantum mechanics,''
  Am.\ J.\ Phys.\ \textbf{68}, 278--281 (2000).

\bibitem{teVrugt2020}
  M.~te Vrugt and R.~Wittkowski,
  ``Projection operators in statistical mechanics: a
  pedagogical approach,''
  Eur.\ J.\ Phys.\ \textbf{41}, 045101 (2020).

\bibitem{Datta1995}
S.~Datta,
\textit{Electronic Transport in Mesoscopic Systems}
(Cambridge University Press, Cambridge, 1995).

\bibitem{Misra1977}
  B.~Misra and E.~C.~G.~Sudarshan,
  J.~Math.\ Phys.\ \textbf{18}, 756 (1977).

\bibitem{Facchi2002}
  P.~Facchi and S.~Pascazio,
  Phys.\ Rev.\ Lett.\ \textbf{89}, 080401 (2002).

\bibitem{FacchiPascazio2008}
  \new{P.~Facchi and S.~Pascazio,
  ``Quantum Zeno dynamics: mathematical and physical aspects,''
  J.~Phys.\ A: Math.\ Theor.\ \textbf{41}, 493001 (2008).}

\bibitem{Itano1990}
  \new{W.~M.~Itano, D.~J.~Heinzen, J.~J.~Bollinger, and
  D.~J.~Wineland,
  ``Quantum Zeno effect,''
  Phys.\ Rev.\ A \textbf{41}, 2295--2300 (1990).}

\bibitem{Lapidus1987}
  I.~R.~Lapidus,
  ``Particle in a square well with a $\delta$-function
  perturbation,''
  Am.\ J.\ Phys.\ \textbf{55}, 172--174 (1987).

\bibitem{Joglekar2009}
  Y.~N.~Joglekar,
  ``Particle in a box with a $\delta$-function potential:
  Strong and weak coupling limits,''
  Am.\ J.\ Phys.\ \textbf{77}, 734--736 (2009).

\bibitem{PedramVahabi2010}
  P.~Pedram and M.~Vahabi,
  ``Exact solutions of a particle in a box with a delta
  function potential: The factorization method,''
  Am.\ J.\ Phys.\ \textbf{78}, 839--841 (2010).

\bibitem{Fernandez2021}
  F.~M.~Fern\'andez,
  ``Variational approach to the Schr\"{o}dinger equation
  with a delta-function potential,''
  Eur.\ J.\ Phys.\ \textbf{43}, 025401 (2022).

\bibitem{RamosdeCastro2024}
  L.~G.~M.~Ramos and A.~S.~de Castro,
  ``Revisiting the bound states of a confined delta
  potential,''
  Eur.\ J.\ Phys.\ \textbf{45}, 035401 (2024),
  DOI: 10.1088/1361-6404/ad2ec1.

\bibitem{Herrera2022}
  W.~J.~Herrera, H.~Vinck-Posada, and S.~G\'omez P\'aez,
  ``Green's functions in quantum mechanics courses,''
  Am.\ J.\ Phys.\ \textbf{90}, 763--769 (2022).

\bibitem{AtkinsonCrater1975}
  D.~A.~Atkinson and H.~W.~Crater,
  ``An exact treatment of the Dirac delta function
  potential in the Schr\"{o}dinger equation,''
  Am.\ J.\ Phys.\ \textbf{43}, 301--304 (1975).

\bibitem{ErmanGadellaUncu2018}
  F.~Erman, M.~Gadella, and H.~Uncu,
  ``On scattering from the one-dimensional multiple Dirac
  delta potentials,''
  Eur.\ J.\ Phys.\ \textbf{39}, 035403 (2018).

\bibitem{Fano1961}
  U.~Fano, Phys.\ Rev.\ \textbf{124}, 1866 (1961).

\bibitem{Miroshnichenko2010}
  A.~E.~Miroshnichenko, S.~Flach, and Y.~S.~Kivshar,
  Rev.\ Mod.\ Phys.\ \textbf{82}, 2257 (2010).

\bibitem{Anderson1961}
  P.~W.~Anderson, Phys.\ Rev.\ \textbf{124}, 41 (1961).

\bibitem{Hewson1993}
  A.~C.~Hewson, \textit{The Kondo Problem to Heavy Fermions},
  Cambridge University Press (1993).

\bibitem{CohenTannoudji1992}
  C.~Cohen-Tannoudji, J.~Dupont-Roc, and G.~Grynberg,
  \textit{Atom--Photon Interactions}, Wiley-Interscience (1992).

\bibitem{Haroche2006}
  S.~Haroche and J.-M.~Raimond,
  \textit{Exploring the Quantum},
  Oxford University Press (2006).

\bibitem{Mahan2000}
  G.~D.~Mahan, \textit{Many-Particle Physics}, 3rd ed.,
  Kluwer Academic / Plenum (2000).

\bibitem{Bruus2004}
  H.~Bruus and K.~Flensberg,
  \textit{Introduction to Many-Body Quantum Theory},
  Oxford University Press (2004).

\bibitem{Taron2013}
  J.~Taron,
  ``Feshbach resonance: A one dimensional example,''
  Am.\ J.\ Phys.\ \textbf{81}, 603--609 (2013).

\bibitem{BreuerPetruccione}
  H.-P. Breuer and F.~Petruccione,
  \textit{The Theory of Open Quantum Systems},
  Oxford University Press (2002).

\bibitem{Daley2014}
  A.~J. Daley, Adv.\ Phys.\ \textbf{63}, 77 (2014).

\bibitem{Kato1966}
  T.~Kato, \textit{Perturbation Theory for Linear Operators}
  (Springer, Berlin, 1966).

\bibitem{BonneauFarautValent2001}
  G.~Bonneau, J.~Faraut, and G.~Valent,
  ``Self-adjoint extensions of operators and the teaching
  of quantum mechanics,''
  Am.\ J.\ Phys.\ \textbf{69}, 322--331 (2001).






\end{thebibliography}
\end{document}